\documentclass[aps,amsmath,amssymb,
reprint,
superscriptaddress,
10pt
]{revtex4-2}

\usepackage{graphicx}
\usepackage{svg}
\usepackage{bm}
\usepackage{siunitx}
\usepackage{xcolor}
\usepackage{xspace}
\usepackage{chemformula}
\usepackage{gensymb}
\usepackage{appendix}

\newcommand{\da}{$\left[111\right]$\xspace}
\newcommand{\db}{$\left[1\overline{1}\overline{1}\right]$\xspace}
\newcommand{\dc}{$\left[\overline{1}1\overline{1}\right]$\xspace}
\newcommand{\dd}{$\left[1\overline{1}\overline{1}\right]$\xspace}

\newcommand{\affil}{HFML-FELIX, Radboud University, Toernooiveld 7, 6525 ED Nijmegen, The Netherlands}

\begin{document}

\title{Ferroelastic domain switching by ultrafast photoinduced strain}

\author{D. G. Lourens}
\email{daniel.lourens@ru.nl}
\affiliation{\affil}
\author{M. Kwaaitaal}
\affiliation{\affil}
\author{C. S. Davies}
\affiliation{\affil}
\author{A. Kirilyuk}
\email{andrei.kirilyuk@ru.nl}
\affiliation{\affil}
\date{\today}

\begin{abstract}
The crystal lattice underpins the fundamental properties of condensed matter, with ferroic order emerging sensitively from atomic coordination. Manipulating the lattice therefore promises a direct way to switch ferroics between their states. Infrared excitation provides an efficient pathway to drive lattice motion and generate transient crystal deformations and strains. Ferroelastics constitute a uniquely direct platform for such control, their order parameter being strain itself. However, whether ultrafast laser-induced lattice distortions can switch ferroelastic order remains unknown. Here we use ultrafast pump–probe microscopy to demonstrate that a single high-amplitude infrared pulse induces strain within the first nanosecond, followed by ferroelastic domain switching several nanoseconds later. The spatial distribution follows that expected from the photoinduced strain field, while their temporal evolution is closely coupled to the strain dynamics. Our observations identify transient strain as the driving field for ferroelastic switching and suggest a general lattice-mediated pathway for controlling ferroic order using infrared light.
\end{abstract}

\maketitle

\section{Introduction}
The properties of non-volatility and hysteresis critically form the basis of numerous modern technologies ranging from memories and control systems to electrical machines and shock absorbers. Ferroelasticity represents the most common manifestation of this nonlinear effect, with spontaneous strain often appearing in tandem with, and even being responsible for defining, ferromagnetic and ferroelectric order in crystals \cite{Nagarajan2002, Li2011}. In general, ferroelastic ordering produces twin domain walls with thicknesses approaching the unit cell level. Such extremely localized topological `defects' can be combined with properties such as ferroelectric polarity, electrical conductivity or even superconductivity, rendering these domain walls a new and exciting state of matter \cite{Marais1991}.  Strain is now widely recognized as playing a very important role in a wide variety of order-disorder structural phase transitions in crystals \cite{Salje1985, Salje1990, Marais1991}. Nevertheless, the desired device properties are related mainly to ferroelectric and ferromagnetic switching. This can be explained by the difficulties of manipulating the ferroelastic order on the required spatial and temporal scales \cite{Gao2014, Li2024}.

It has long been known that all-optical switching of domains offers an intriguing and technologically compelling pathway to directly manipulate the order parameter defining a multitude of magnetic and ferroelectric materials \cite{Subedi2014, RubioMarcos2017, Vats2019, Wang2019, Stupakiewicz2021, Janssen2023, Gidding2024, Kwaaitaal2024, Davies2024, Zeng2025}. Several mechanisms have been discovered to be responsible for switching in these materials, including thermal-, electronic-, opto-magnetic, and phonon-mediated ones \cite{Guo2021}. Recent works have highlighted the potential of using photo-induced strains to modify the crystal potential, resulting in a switching of the corresponding ferroic parameter \cite{Stupakiewicz2021,Kwaaitaal2024, Janssen2023}. So far, however, no all-optical switching has been reported in ferroelastic crystals, despite these materials plausibly respresenting an ideal system for studying strain-driven switching mechanics due to their direct coupling of strain with ferroelastic order. In contrast, strain in different types of ferroics only indirectly influences the order parameter (e.g. via the piezoelectric effect in ferroelectrics \cite{Kwaaitaal2024}).

Here, we demonstrate that picosecond optical pulses in the mid-infrared spectral range can all-optically switch ferroelastic domains in lanthanum aluminate (\ch{LaAlO3}). We directly observe that the infrared excitation generates strain within nanoseconds, with the switching following just several nanoseconds later. \ch{LaAlO3} has been widely used as a compatible substrate for a large variety of thin films, ranging from high-temperature superconducting materials \cite{Simon1989} to magnetic and ferroelectric thin films \cite{Rambabu2023}. Its basic properties, in particular possible domain orientations and infrared optical properties across phononic resonances have been studied extensively \cite{Michael1992, Hayward2002, Vermeulen2016, Elias2018, Rizwan2019}, which makes it an ideal candidate for investigating the coupling between strain and ferroelastic order possibly leading to all-optical domain switching. 

\section{Results}
\ch{LaAlO3} is a perovskite with a paraelastic cubic phase above its Curie temperature ($T_C\approx800$ K \cite{Vermeulen2016, Hayward2005}) and ferroelastic rhombohedral phase below the Curie temperature \cite{Lehnert2000}.
The rhombohedral crystal structure of ferroelastic \ch{LaAlO3} has lattice parameters of unit cell length $a=5.3 \text{\r{A}}$ and angle $\alpha=90.1$° \cite{Lehnert2000, Hayward2005} and results in four possible domain orientations, denoted as \da, \db, \dc and \dd\ (Fig. \ref{fig:Switching-strain}a). This crystal structure of individual domains requires domain wall orientations that minimize interface strain between adjacent domains \cite{Vermeulen2016}. Table \ref{tab:boundary-orientations} shows such allowed orientations of domain walls. 

\begin{table}[]
    \caption{\textbf{Domain wall orientations in \ch{LaAlO3}.} Domain orientations (square brackets) given by their optical axis and their non-strained domain boundary orientation (round brackets) in rhombohedral \ch{LaAlO3}.}
    \label{tab:boundary-orientations}
    \centering
    {\renewcommand{\arraystretch}{1.5}
    \begin{tabular}{c | c c c c}
         & \da & \db & \dc & \dd \\
    \hline
        \da & - & $(011)$ & $(010)$ & $(001)$ \\ 
        \db &   &    -    & $(001)$ & $(010)$ \\ 
        \dc &   &         &    -    & $(01\overline{1})$ \\
        \dd &   &         &         & -
    \end{tabular}
    }
\end{table}

Domain dynamics were induced with mid- to far-infrared pulses of a free-electron laser (FEL), tuned to frequencies where the dielectric constant values approaches zero, the so-called  Epsilon-Near-Zero (ENZ) regime. Previous research has shown that it is this frequency range that results in strong light-matter interaction and subsequent domain switching in ferrimagnets \cite{Stupakiewicz2021} and ferroelectrics \cite{Kwaaitaal2024}. Detection was achieved through a single-shot pump-probe scheme, using femtosecond pulses in the visible part of spectrum as a probe.

Due to the strong birefringence in \ch{LaAlO3} \cite{Hayward2005}, polarization microscopy was chosen for the detection of both domain switching and strain dynamics. The full polarization information was captured using a polarization sensitive camera. Data analysis was performed on the resulting intensity (I), Angle of Polarization (AoP) and Degree of Linear Polarization (DoLP), with the DoLP being linearly proportional to the induced strain in this regime (see Eq. \eqref{eq:strain-birefringence} in the Methods).

\subsection{Domain switching following IR excitation}
In our experiments, a single-domain $\langle100\rangle$ \ch{LaAlO3} crystal was used. The sample was pumped by picosecond-long far-infrared pulses (either single or coming in short trains) focussed with an off-axis parabolic mirror to a spatially Gaussian spot with a diameter of 100-200 µm depending on the pump wavelength. The resulting dynamics are recorded in a pump-probe scheme using full-field polarization microscopy with a 520 nm laser as a probe. After excitation with the pump pulse, both transient strain and domain switching were observed with characteristic lifetimes of a few hundred microseconds. 

\begin{figure}
    \centering
    \includegraphics[width=3.3in]{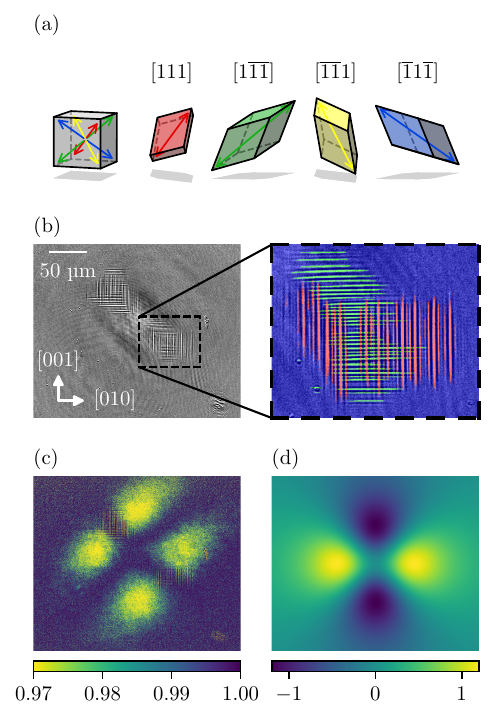}
    \caption{
    \textbf{Domain structure and strain in IR-pumped \ch{LaAlO3}.}
    (a) Schematic representation of the stretching of a diagonal from the paraelastic cubic phase to the ferroelastic rhombohedral phase in \ch{LaAlO3}. The four domain directions are color-coded blue, red, green, and yellow.
    (b) Differential map of the measured intensity 30 µs after excitation with a far-infrared burst of pulses. Two areas with switched domains (horizontal and vertical stripes) appear. A part of the pattern is colored to show its domain orientation. The colors correspond to the domain orientations as shown in (a).
    (c) Measured degree of linear polarization (DoLP). The intensity of the four lobes corresponds to the laser-induced strain.
    (d) Calculated (normalized) rotation of the optical axis due to a Gaussian strain profile.
    }
    \label{fig:Switching-strain}
\end{figure}

An example of the resulting domain structure is shown in Figure \ref{fig:Switching-strain}b. Domains appear as both horizontal and vertical stripes on the edges of the excitation spot, along the $\left[01\overline{1}\right]$ crystal diagonal, with no domains appearing in the center of the excitation spot or along the $\left[011\right]$ diagonal. In the magnified view of Figure \ref{fig:Switching-strain}b, part of the domain pattern is colored to highlight the different domain states. The initial domain state is colored blue, and the two switched domain states are colored red and green, corresponding to the colors shown in Figure \ref{fig:Switching-strain}a. 

We determined the optical axis of our sample before the excitation to be \dc using angle-dependent birefringence measurements described by Glazer et al.\ \cite{Glazer1996}. With Table \ref{tab:boundary-orientations} we can conclude that formed domains have either \da or \db orientations.

To explain these results, we apply a theory for strain-driven domain switching. Such theories have been previously used to describe the all-optical switching mechanism in several ferroics \cite{Stupakiewicz2021, Kwaaitaal2024}, in which a spatially Gaussian laser pulse creates a Gaussian strain profile in the crystal. In the case of the ferroelectric \ch{BaTiO3} for example, this strain profile induces a piezoelectric displacement field that aids the domain switching. In the case of the ferroelastic \ch{LaAlO3}, such an induced strain profile can directly couple to switching without the need for an intermediate effect.

By inducing a strain in a birefringent crystal such as \ch{LaAlO3}, the orientation of its optical axis can be modified \cite{Kwaaitaal2024a}. The effect of a 2-dimensional Gaussian strain profile on the angle of the optical axis is calculated in the Supplementary Material, and shown in Figure \ref{fig:Switching-strain}d. The initial angle of the optical axis is taken along the $\left[011\right]$ diagonal, which corresponds to the sample orientation in our experiments. A four-lobed pattern emerges, showing both a positive and negative rotation with respect to the initial angle of the optical axis.

In our experiments, the polarization of the probe laser is aligned along the optical axis of the crystal ($\left[0\overline{1}1\right]$ direction). In the absence of a pump pulse, this results in no change in polarization and thus the measured degree of linear polarization (DoLP) is 100\%. Pumping the crystal can produce strains that result in a change in the dielectric permeability tensor $\eta_{ij}=\eta^0_{ij}+\Delta\eta_{ij}$ and thus a rotation of the optical axis, which induces birefringence \cite{Slezak2020}. In our experiments the rotation of the optical axis is sufficiently small, giving the relation $\mathrm{DoLP}\approx1-2\left(\vartheta\sin{\Delta\varphi}\right)^2$, where $\vartheta$ is the rotation of the optical axis, and $\Delta\varphi$ is the phase shift between the ordinary and extraordinary axis. Figure \ref{fig:Switching-strain}d shows the measured DoLP after an excitation with a pump pulse. The agreement between the calculated rotation of optical axis and the measured resulting DoLP confirms that the measured change in polarization is produced by a Gaussian-shaped strain profile.

Thermoelastic simulations were performed to calculate the laser-induced thermal strains in \ch{LaAlO3}. These were carried out assuming an uncoupled theory, in which it is assumed that induced strains have negligible influence on the heat flow in the crystal. A detailed description of the simulations can be found in the Supplementary Material. A \qty{1}{mJ} pump with a FWHM of \qty{110}{\um}, and an absorption coefficient of \qty{0.03}{\um^{-1}} were chosen as parameters for the simulation, matching closely with the experimental parameters for a macropulse with a central wavelength of \qty{12}{\um}. The calculated resulting strain is shown in the Supplementary Material. The calculated values are on the order of \num{d-3}, which is close to previously measured spontaneous strain \cite{Hayward2002}.

\begin{figure}
    \centering
    \includegraphics[width=2.5in]{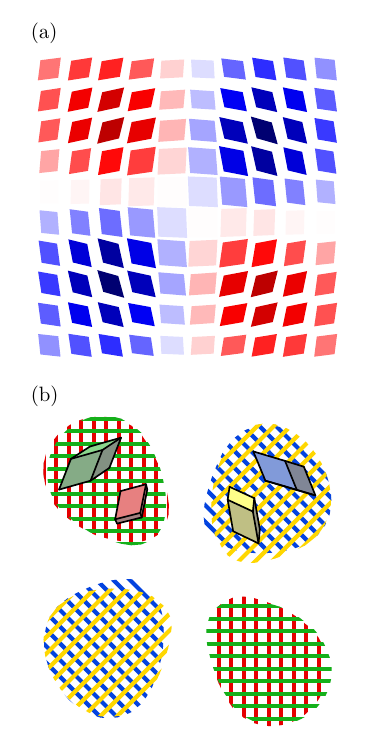}
    \caption{\textbf{Spatial distribution (\qtyproduct{300 x 300}{\um}) of laser-induced body diagonal stretching.} (a) Two-dimensional deformation of a square lattice when subjected to the calculated surface strain profile. Each parallellogram shows the (exaggerated) deformation The color shows the (normalized) shear strain in every parallellogram. (b) Striped areas show which diagonals are stretched relative to the shortest diagonal, with the rhombuses corresponding to those in Fig. \ref{fig:Switching-strain}(a). Similar stretches are found for the \da and \db diagonals, as well as for the \dc and \dd diagonals.}
    \label{fig:DomainPreference}
\end{figure}

The strain at any point may be visualized by displaying the (greatly exaggerated) distortion of a square subjected to the strain. The resulting deformation in the $(100)$ plane is displayed in Fig. \ref{fig:DomainPreference}(a). The distorted unit cells have a diamond-like shape with a $\left[011\right]$ stretch on the negative diagonal and a $\left[0\overline{1}1\right]$ stretch on the positive diagonal.

A simple model was used to calculate which domain states are favoured by to the induced strain. For each point in the sample, the lengths of the four body diagonals were compared. For each diagonal, this point is coloured when the ratio between the diagonal and the shortest diagonal reaches above a certain threshold. Since the strains are largely located near the surface of the crystal, a 2D projection was made for simpler interpretation. The results are shown in Fig. \ref{fig:DomainPreference}(b). The \da and \db diagonals as well as the \dc and \dd diagonals show similar lengths, resulting in very minor preferences between the corresponding domain states.

To relate these findings to our experimental results (Fig. \ref{fig:Switching-strain}b), we note that the crystal initially consists of a single-domain state that is represented by strain on the positive diagonal (\dc-stretch). Since the thermal strain produces comparable stretches on the \dc and \dd diagonals, it produces no meaningful driving force for switching towards the \dd domain state. In contrast, the \da and \db stretches on the negative diagonal are substantially larger than the stretch on the \dc diagonal, thereby providing a driving force for switching toward the \da and \db domain orientations. These results thus suggest that switching only occurs on the negative diagonal, where the strain favours the \da\ and \db\ domain orientations sufficiently over the initial orientation, as indicated by the colors in Fig. \ref{fig:Switching-strain}b. This is in excellent agreement with the experimental observations in Fig. \ref{fig:Switching-strain}b.

\subsection{Temporal evolution of domains}
\begin{figure}
    \centering
    \includegraphics[width=3.2in]{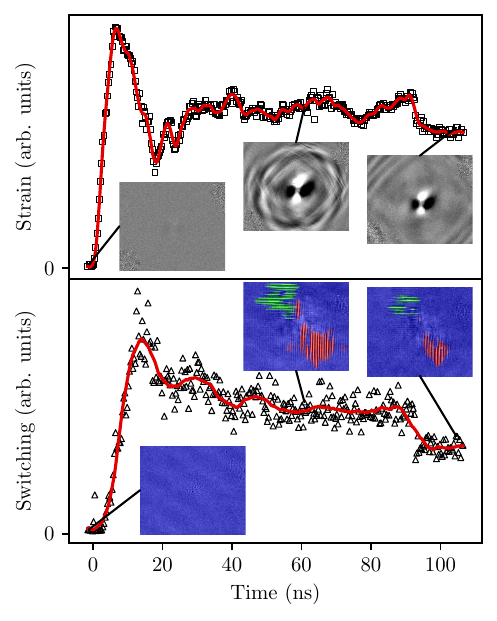}
    \caption{
    \textbf{Temporal evolution of domain and strain dynamics in IR-pumped \ch{LaAlO3}.}
    Induced strain and domain switching after pumping with a single micropulse at 12 µm. The strain graph shows the magnitude of the quasi-static strain, averaged over an area of $1\times1$ µm$^2$. The switching graph shows the total area of switched domains. The images shown for both the strain and switching are taken at -1.2, 60.9 and 104.5 ns, with the switching images zoomed in $(\times5)$ to show better detail.}
    \label{fig:time-dep-switching}
\end{figure}

To further understand the relation of the behaviour of strain with that of switching, we studied their dynamics on the pico- to nanosecond time scale, using single-shot imaging with a polarization-sensitive microscope to observe the transient strain pattern and domains. We used 1-3 ps pulses with wavelengths of 12-14 µm as a pump, and synchronized 400 fs pulses with a wavelength of 515 nm as a probe. Acoustic strain waves and a quasi-static Gaussian strain profile appear within half a nanosecond after the excitation. 
The Gaussian-shaped strain is manifested as a four-lobed pattern and peaks around 6.5 ns before exponentially decaying over several hundred microseconds, while the strain waves propagate outward from the excitation point with the velocities of sound (Fig. \ref{fig:time-dep-switching}).

Domains start to emerge with a delay of about 2 ns after the appearance of the strain pattern, with the switched area peaking around 14 ns before decaying over several hundred microseconds as well. Both the amplitude of strain and the domain size show visible oscillations, on a time scale of tens of nanoseconds. The variations in the measurable strain amplitude is very well explained by the propagating sounds waves mentioned above. Intriguingly, there is a certain degree of correlation between between strain and the size of the domains. While the latter data are more noisy, some features such as the first overshot are clearly correlated. The fact that on this short time scale the reversed domains can oscillate in size underlines their strong coupling with the driving force that is the strain. The domain dynamics are stabilized after the first period of 100 ns, with a subsequent decay over hundreds of microseconds.
While both strain and switching decay on a hundreds of microseconds timescale, the observed decay of the domains is slightly faster than the decay of the strain. This can be explained by the maximum size of the domains, which is not sufficient for a (meta)stable state. As soon as the amplitude of the strain is reduced below the stability threshold (i.e. not sufficiently favouring the switched domain), domains shrink and vanish, driven by crystal thermodynamics.




\section{Discussion}
In this work, we have demonstrated that pumping ferroelectric \ch{LaAlO3} with infrared radiation enables all-optical switching of ferroelastic domains. Switching occurs when the induced strain sufficiently favours a certain domain state over the initial orientation, and first appears several nanoseconds after the initial excitation with the pump-pulse. We have shown how induced strain profiles form a direct pathway towards ferroelastic domain switching. Thermoelastic simulations show laser heating as a promising candidate for the generation of these strain profiles, with induced strain values approaching those of spontaneous strain. To further validate this mechanism for inducing strain, more sophisticated thermoelastic models could be employed to obtain a more accurate description of the thermally induced strain. These predictions could be compared with x-ray diffraction measurements to determine whether heating alone sufficiently explains the observed strain generation \cite{RosePetruck1999, Whiteley2019, Mariette2021, Wang2022}.

The proposed method of domain switching opens the possibility of precise domain engineering using geometric masking of the pump pulse to create desired strain patterns, and should work for different ferroelastics as well, as long as the induced strain is engineered correctly. Similar models of strain-driven switching of the ferroic order have been reported in ferrimagnetic iron garnets \cite{Stupakiewicz2021} and in the ferroelectric \ch{BaTiO3} \cite{Kwaaitaal2024}, highlighting the universality of this switching pathway.

\section*{Acknowledgements}
The authors thank all technical staff of the FELIX facility for
technical support.
D.G.L. acknowledges funding by the Max Planck–Radboud University Center for Infrared Free Electron Laser Spectroscopy.
C.S.D. acknowledges support from the European Research Council ERC Grant Agreement No. 101115234 (HandShake), and A.K. acknowledges support from the European Research Council ERC Grant Agreement No. 101141740 (INTERPHON).

\section*{Methods}
\subsection{Materials}
The \ch{LaAlO3} crystal used in this work was made by Biotain Crystal Co., Limited and is a single-domain $\langle100\rangle$ crystal of $5\times5\times0.5$ mm with both sides polished (Ra $<0.5$ nm).

\subsection{Pumping with IR-FELs}
For this work, the radiation of two of the free-electron lasers available in the FELIX facility in Nijmegen, the Netherlands \textemdash named FEL-1 and FELICE \textemdash was used as a pump to induce domain switching in the \ch{LaAlO3} crystal. Both lasers deliver transform limited pulses of infrared light with a bandwidth of 0.5-1\%. The cavity-dump setup that was designed for the FELICE FEL delivers picosecond pulses with a repetition rate of 5 Hz and energies up to 130 µJ \cite{Janssen2022}, and was used with a wavelength of 12 µm for the time-resolved measurements discussed above. The FEL-1 laser delivers 10-µs-long bursts of picosecond pulses with a repetition rate of 25 MHz (macropulses) with energies up to 4 mJ and was used in a wavelength range of 8-100 µm to reach higher pump powers and enable the induction of larger strains.

\subsection{Polarizing microscopy as a probe}
The crystal was subsequently probed in a pump-probe full-field imaging scheme. For the time-dependent measurements with a picosecond resolution, pump pulses from FELICE were combined with optical pulses delivered by a tabletop continuously-amplified Yb-doped fiber laser (Tangor 100, AMPLITUDE). The tabletop laser delivers 400-fs-long pulses at a central wavelength of 515 nm and a repetition rate of 5 Hz, with single pulses having an energy of up to 0.5 mJ. Pump pulses from FEL-1 were combined with a continuous laser (MatchBox) with a central wavelength of 520 nm as a probe.

Light of the probe lasers was gathered on a polarizing camera with a 27 µs exposure time (Thorlabs Kiralux CMOS CS505MUP1). This camera has an array of wiregrid polarizers, with each pixel having the transmission axis either 0°, 45°, -45°, or 90°, in a repeating pattern. This creates $2\times2$ superpixels which combined capture the full polarization information of the probe lasers. From this data, the intensity and degree of linear polarization (DoLP) were calculated using Stokes parameters. The intensity picture is used to measure domain switching, while the DoLP gives a measure of induced strain in the crystal \cite{Kwaaitaal2024a}. The polarization of the probe laser was aligned parallel to the optical axis of the crystal ($\left[0\overline{1}1\right]$ orientation), so that without an external strain the DoLP will be 100\%. After excitation with the pump, the induced strain will cause the optical axis of the crystal to rotate, with a stronger strain inducing a larger rotation of the optical axis. This rotation induces a birefringence, making the probe light elliptical and decreasing the DoLP.

The rotation of the optical axis as a function of a Gaussian strain profile can be calculated as follows, by considering the change in the relative dielectric impermeability tensor $\eta=1/n^2$ \cite{Slezak2020}:

\begin{equation}
    \Delta \eta_{ij} = p_{ijkl}\varepsilon_{kl},
\end{equation}

where $p_{ijkl}$ are the elasto-optic coefficients of the material, and $\varepsilon_{kl}$ the strain components
\begin{align}
\label{eq:strainxx}
    S_{xx} &\propto \frac{1}{r^2} \left\{ 
                \left( 1 - e^{-\frac{r^2}{2\sigma^2}} \right)
                \left( 1 - \frac{2x^2}{r^2} \right)
                +\frac{x^2}{\sigma^2}e^{-\frac{r^2}{2\sigma^2}}
            \right\} \\
\label{eq:strainyy}
    S_{yy} &\propto \frac{1}{r^2} \left\{ 
                \left( 1 - e^{-\frac{r^2}{2\sigma^2}} \right)
                \left( 1 - \frac{2y^2}{r^2} \right)
                +\frac{y^2}{\sigma^2}e^{-\frac{r^2}{2\sigma^2}}
            \right\} \\
\label{eq:strainxy}
    S_{xy} &\propto \frac{xy}{r^2} \left\{ 
                \frac{2}{r^2}
                 \left( 1 - e^{-\frac{r^2}{2\sigma^2}} \right)
                 -\frac{1}{\sigma^2}e^{-\frac{r^2}{2\sigma^2}}
            \right\}.
\end{align}
The full dielectric impermeability is then given by
$\eta_{ij} = \eta_{ij}^0 + \Delta \eta_{ij}$, with the components
\begin{equation}
    \begin{aligned}
        \eta_{xy} &= \beta\cdot p_{xyxy} \varepsilon_{xy} \\
        \eta_{xx} &= \frac{1}{n_{xx}^2} + \beta\cdot\left( p_{xxxx}\varepsilon_{xx} + p_{xxyy}\varepsilon_{yy} \right) \\
        \eta_{yy} &= \frac{1}{n_{yy}^2} + \beta\cdot\left( p_{yyxx}\varepsilon_{xx} + p_{yyyy}\varepsilon_{yy} \right).
    \end{aligned}
\end{equation}
Linearizing the matrix $\eta$ gives an equation for the angle of rotation of the optical axis
\begin{equation}
    \begin{aligned}
        \tan(2\alpha) &= \frac{2\eta_{xy}}{\eta_{yy} - \eta_{xx}} \\
         &= \frac{\beta\cdot2p_{xyxy} \varepsilon_{xy}}{
            \eta_0
            + \beta\cdot\left(
                p_{X}\varepsilon_{xx}
                + p_{Y}\varepsilon_{yy}
            \right)
            }.
    \end{aligned}
\end{equation}
where $\beta$ is the amplitude of the Gaussian strain profile, and
\begin{equation}
    \begin{aligned}
        \eta_{0} &= \frac{1}{n_{yy}^2} - \frac{1}{n_{xx}^2} \\
        p_{X} &= p_{yyxx} - p_{xxxx} \\
        p_{Y} &= p_{yyyy} - p_{xxyy}.
    \end{aligned}
\end{equation}
In the case where the strain is sufficiently small, thus $\beta \ll \eta_{0}$:
\begin{equation}
\label{eq:strain-birefringence}
    \begin{aligned}
        \alpha &\approx \frac{1}{2}\arctan{\left( 
        \frac{\beta}{\eta_{0}}\cdot 2p_{xyxy} \varepsilon_{xy}
        \right)} \\
        &\approx \frac{\beta}{\eta_{0}}\cdot p_{xyxy} \varepsilon_{xy},
    \end{aligned}
\end{equation}
showing that the angle of rotation is proportional to the shear components of the strain profile.

\section*{Author declarations}
\subsection*{Conflict of Interest}
The authors have no conflicts to disclose.

\section*{Data availability}
The data that support the findings of this study are available from the corresponding author upon reasonable request.

\bibliography{bibliography_arXiv.bib}

@InBook{Simon1989,
  author    = {Simon, R. W. and Lee, A. E. and Platt, C. E. and Daly, K. P. and Luine, J. A. and Eom, C. B. and Rosenthal, P. A. and Wu, X. D. and Venkatesan, T.},
  pages     = {337--346},
  publisher = {Springer US},
  title     = {Growth of High-Temperature Superconductor Thin Films on Lanthanum Aluminate Substrates},
  year      = {1989},
  isbn      = {9781468456585},
  booktitle = {Science and Technology of Thin Film Superconductors},
  doi       = {10.1007/978-1-4684-5658-5_40},
}

@Article{Rambabu2023,
  author    = {Rambabu, A. and Sundaresan, A},
  journal   = {Materials Today: Proceedings},
  title     = {Magnetic and electrical properties of LaCoO3 - LaNiO3 epitaxial thin films on LaAlO3 substrate},
  year      = {2023},
  issn      = {2214-7853},
  month     = mar,
  doi       = {10.1016/j.matpr.2023.02.275},
  publisher = {Elsevier BV},
}

@Article{Rizwan2019,
  author    = {Rizwan, Muhammad and Gul, Samina and Iqbal, Tahir and Mushtaq, Uzma and Farooq, M Hassan and Farman, Muhammad and Bibi, Rabia and Ijaz, Mohsin},
  journal   = {Materials Research Express},
  title     = {A review on perovskite lanthanum aluminate (LaAlO3), its properties and applications},
  year      = {2019},
  issn      = {2053-1591},
  month     = sep,
  number    = {11},
  pages     = {112001},
  volume    = {6},
  doi       = {10.1088/2053-1591/ab4629},
  publisher = {IOP Publishing},
}

@Article{Michael1992,
  author    = {Michael, Peter C. and Trefny, John U. and Yarar, Baki},
  journal   = {Journal of Applied Physics},
  title     = {Thermal transport properties of single crystal lanthanum aluminate},
  year      = {1992},
  issn      = {1089-7550},
  month     = jul,
  number    = {1},
  pages     = {107--109},
  volume    = {72},
  doi       = {10.1063/1.352166},
  publisher = {AIP Publishing},
}

@Article{Elias2018,
  author    = {Elias, Badal H and Ilyas, Bahaa M and Saadi, Nawzat S},
  journal   = {Materials Research Express},
  title     = {A first principle study of the perovskite lanthanum aluminate},
  year      = {2018},
  issn      = {2053-1591},
  month     = jul,
  number    = {8},
  pages     = {086302},
  volume    = {5},
  doi       = {10.1088/2053-1591/aad15f},
  publisher = {IOP Publishing},
}

@Article{Hayward2002,
  author    = {Hayward, S A and Redfern, S A T and Salje, E K H},
  journal   = {Journal of Physics: Condensed Matter},
  title     = {Order parameter saturation in LaAlO3},
  year      = {2002},
  issn      = {0953-8984},
  month     = oct,
  number    = {43},
  pages     = {10131--10144},
  volume    = {14},
  doi       = {10.1088/0953-8984/14/43/311},
  publisher = {IOP Publishing},
}

@Article{Hayward2005,
  author    = {Hayward, S. A. and Morrison, F. D. and Redfern, S. A. T. and Salje, E. K. H. and Scott, J. F. and Knight, K. S. and Tarantino, S. and Glazer, A. M. and Shuvaeva, V. and Daniel, P. and Zhang, M. and Carpenter, M. A.},
  journal   = {Physical Review B},
  title     = {Transformation processes inLaAlO3: Neutron diffraction, dielectric, thermal, optical, and Raman studies},
  year      = {2005},
  issn      = {1550-235X},
  month     = aug,
  number    = {5},
  pages     = {054110},
  volume    = {72},
  doi       = {10.1103/physrevb.72.054110},
  publisher = {American Physical Society (APS)},
}

@Article{Kwaaitaal2024,
  author    = {Kwaaitaal, M. and Lourens, D. G. and Davies, C. S. and Kirilyuk, A.},
  journal   = {Nature Photonics},
  title     = {Epsilon-near-zero regime enables permanent ultrafast all-optical reversal of ferroelectric polarization},
  year      = {2024},
  issn      = {1749-4893},
  month     = apr,
  doi       = {10.1038/s41566-024-01420-3},
  publisher = {Springer Science and Business Media LLC},
}

@Article{Kwaaitaal2024a,
  author    = {Kwaaitaal, Maarten and Lourens, Daniel G. and Davies, Carl S. and Kirilyuk, Andrei},
  journal   = {Scientific Reports},
  title     = {Disentangling thermal birefringence and strain in the all-optical switching of ferroelectric polarization},
  year      = {2024},
  issn      = {2045-2322},
  month     = oct,
  number    = {1},
  volume    = {14},
  doi       = {10.1038/s41598-024-75670-0},
  publisher = {Springer Science and Business Media LLC},
}

@Article{Stupakiewicz2021,
  author     = {A. Stupakiewicz and C. S. Davies and K. Szerenos and D. Afanasiev and K. S. Rabinovich and A. V. Boris and A. Caviglia and A. V. Kimel and A. Kirilyuk},
  journal    = {Nature Physics},
  title      = {Ultrafast phononic switching of magnetization},
  year       = {2021},
  month      = {jan},
  number     = {4},
  pages      = {489--492},
  volume     = {17},
  comment    = {Theory of strain-induced switching.},
  doi        = {10.1038/s41567-020-01124-9},
  publisher  = {Springer Science and Business Media {LLC}},
  ranking    = {rank5},
  readstatus = {skimmed},
}

@Article{Lehnert2000,
  author    = {Lehnert, H. and Boysen, Hans and Dreier, P. and Yu, Y.},
  journal   = {Zeitschrift für Kristallographie - Crystalline Materials},
  title     = {Room temperature structure of LaAlO3},
  year      = {2000},
  issn      = {2194-4946},
  month     = mar,
  number    = {3},
  pages     = {145--147},
  volume    = {215},
  doi       = {10.1524/zkri.2000.215.3.145},
  publisher = {Walter de Gruyter GmbH},
}

@Article{Vermeulen2016,
  author    = {Vermeulen, Paul A. and Kumar, Anil and ten Brink, Gert H. and Blake, Graeme R. and Kooi, Bart J.},
  journal   = {Crystal Growth \& Design},
  title     = {Unravelling the Domain Structures in GeTe and LaAlO3},
  year      = {2016},
  issn      = {1528-7505},
  month     = sep,
  number    = {10},
  pages     = {5915--5922},
  volume    = {16},
  doi       = {10.1021/acs.cgd.6b00960},
  publisher = {American Chemical Society (ACS)},
}

@Article{Janssen2022,
  author    = {T. Janssen and C. S. Davies and M. Gidding and V. Chernyy and J. M. Bakker and A. Kirilyuk},
  journal   = {Review of Scientific Instruments},
  title     = {Cavity-dumping a single infrared pulse from a free-electron laser for two-color pump{\textendash}probe experiments},
  year      = {2022},
  month     = {apr},
  number    = {4},
  pages     = {043007},
  volume    = {93},
  doi       = {10.1063/5.0081862},
  publisher = {{AIP} Publishing},
}

@Article{Gidding2024,
  author    = {Gidding, Maxime and Davies, Carl S. and Kirilyuk, Andrei},
  journal   = {Physical Review B},
  title     = {Reorientation of magnetic stripe domains by mid-infrared pulses},
  year      = {2024},
  issn      = {2469-9969},
  month     = feb,
  number    = {6},
  pages     = {l060408},
  volume    = {109},
  doi       = {10.1103/physrevb.109.l060408},
  publisher = {American Physical Society (APS)},
}

@Article{Davies2024,
  author    = {Davies, C. S. and Fennema, F. G. N. and Tsukamoto, A. and Razdolski, I. and Kimel, A. V. and Kirilyuk, A.},
  journal   = {Nature},
  title     = {Phononic switching of magnetization by the ultrafast Barnett effect},
  year      = {2024},
  issn      = {1476-4687},
  month     = apr,
  number    = {8008},
  pages     = {540--544},
  volume    = {628},
  doi       = {10.1038/s41586-024-07200-x},
  publisher = {Springer Science and Business Media LLC},
}

@Article{Janssen2023,
  author    = {Janssen, T. and Gidding, M. and Davies, C. S. and Kimel, A. V. and Kirilyuk, A.},
  journal   = {Physical Review B},
  title     = {Strain-induced magnetic pattern formation in antiferromagnetic iron borate},
  year      = {2023},
  issn      = {2469-9969},
  month     = oct,
  number    = {14},
  pages     = {l140405},
  volume    = {108},
  doi       = {10.1103/physrevb.108.l140405},
  publisher = {American Physical Society (APS)},
}

@Article{Glazer1996,
  author    = {Glazer, A. M. and Lewis, J. G. and Kaminsky, W.},
  journal   = {Proceedings of the Royal Society of London. Series A: Mathematical, Physical and Engineering Sciences},
  title     = {An automatic optical imaging system for birefringence},
  year      = {1996},
  issn      = {1471-2946},
  month     = dec,
  number    = {1955},
  pages     = {2751--2765},
  volume    = {452},
  doi       = {10.1098/rspa.1996.0145},
  publisher = {The Royal Society},
}

@Article{Guo2021,
  author    = {Guo, Jiaxing and Chen, Wenwen and Chen, Haisheng and Zhao, Yanan and Dong, Feng and Liu, Weiwei and Zhang, Yang},
  journal   = {Advanced Optical Materials},
  title     = {Recent Progress in Optical Control of Ferroelectric Polarization},
  year      = {2021},
  issn      = {2195-1071},
  month     = mar,
  number    = {23},
  volume    = {9},
  doi       = {10.1002/adom.202002146},
  publisher = {Wiley},
}

@Article{Marais1991,
  author    = {Marais, Simon and Heine, Volker and Nex, Chris and Salje, Ekhard},
  journal   = {Physical Review Letters},
  title     = {Phenomena due to strain coupling in phase transitions},
  year      = {1991},
  issn      = {0031-9007},
  month     = may,
  number    = {19},
  pages     = {2480--2483},
  volume    = {66},
  doi       = {10.1103/physrevlett.66.2480},
  publisher = {American Physical Society (APS)},
}

@Article{Salje1985,
  author    = {Salje, E.},
  journal   = {Physics and Chemistry of Minerals},
  title     = {Thermodynamics of sodium feldspar I: Order parameter treatment and strain induced coupling effects},
  year      = {1985},
  issn      = {1432-2021},
  number    = {2},
  pages     = {93--98},
  volume    = {12},
  doi       = {10.1007/bf01046833},
  publisher = {Springer Science and Business Media LLC},
}

@Article{Salje1990,
  author    = {Salje, E.},
  journal   = {Ferroelectrics},
  title     = {Phase transitions in ferroelastic and co-elastic crystals},
  year      = {1990},
  issn      = {1563-5112},
  month     = apr,
  number    = {1},
  pages     = {111--120},
  volume    = {104},
  doi       = {10.1080/00150199008223816},
  publisher = {Informa UK Limited},
}

@Article{Nagarajan2002,
  author    = {Nagarajan, V. and Roytburd, A. and Stanishevsky, A. and Prasertchoung, S. and Zhao, T. and Chen, L. and Melngailis, J. and Auciello, O. and Ramesh, R.},
  journal   = {Nature Materials},
  title     = {Dynamics of ferroelastic domains in ferroelectric thin films},
  year      = {2002},
  issn      = {1476-4660},
  month     = dec,
  number    = {1},
  pages     = {43--47},
  volume    = {2},
  doi       = {10.1038/nmat800},
  publisher = {Springer Science and Business Media LLC},
}

@Article{Li2011,
  author    = {Li, L.J. and Lei, C.H. and Shu, Y.C. and Li, J.Y.},
  journal   = {Acta Materialia},
  title     = {Phase-field simulation of magnetoelastic couplings in ferromagnetic shape memory alloys},
  year      = {2011},
  issn      = {1359-6454},
  month     = apr,
  number    = {7},
  pages     = {2648--2655},
  volume    = {59},
  doi       = {10.1016/j.actamat.2011.01.001},
  publisher = {Elsevier BV},
}

@Article{Li2024,
  author    = {Li, Tianyu and Deng, Shiqing and Liu, Hui and Chen, Jun},
  journal   = {Chemical Reviews},
  title     = {Insights into Strain Engineering: From Ferroelectrics to Related Functional Materials and Beyond},
  year      = {2024},
  issn      = {1520-6890},
  month     = may,
  number    = {11},
  pages     = {7045--7105},
  volume    = {124},
  doi       = {10.1021/acs.chemrev.3c00767},
  publisher = {American Chemical Society (ACS)},
}

@Article{Gao2014,
  author    = {Gao, Peng and Britson, Jason and Nelson, Christopher T. and Jokisaari, Jacob R. and Duan, Chen and Trassin, Morgan and Baek, Seung-Hyub and Guo, Hua and Li, Linze and Wang, Yiran and Chu, Ying-Hao and Minor, Andrew M. and Eom, Chang-Beom and Ramesh, Ramamoorthy and Chen, Long-Qing and Pan, Xiaoqing},
  journal   = {Nature Communications},
  title     = {Ferroelastic domain switching dynamics under electrical and mechanical excitations},
  year      = {2014},
  issn      = {2041-1723},
  month     = may,
  number    = {1},
  volume    = {5},
  doi       = {10.1038/ncomms4801},
  publisher = {Springer Science and Business Media LLC},
}

@Article{Subedi2014,
  author    = {Subedi, Alaska and Cavalleri, Andrea and Georges, Antoine},
  journal   = {Physical Review B},
  title     = {Theory of nonlinear phononics for coherent light control of solids},
  year      = {2014},
  issn      = {1550-235X},
  month     = jun,
  number    = {22},
  pages     = {220301},
  volume    = {89},
  doi       = {10.1103/physrevb.89.220301},
  publisher = {American Physical Society (APS)},
}

@Article{Zeng2025,
  author    = {Zeng, Z. and Först, M. and Fechner, M. and Prabhakaran, D. and Radaelli, P. G. and Cavalleri, A.},
  journal   = {Science},
  title     = {Photo-induced nonvolatile rewritable ferroaxial switching},
  year      = {2025},
  issn      = {1095-9203},
  month     = oct,
  number    = {6769},
  pages     = {195--198},
  volume    = {390},
  doi       = {10.1126/science.adz5230},
  publisher = {American Association for the Advancement of Science (AAAS)},
}

@Article{RubioMarcos2017,
  author    = {Rubio-Marcos, Fernando and Ochoa, Diego A. and Del Campo, Adolfo and García, Miguel A. and Castro, Germán R. and Fernández, José F. and García, José E.},
  journal   = {Nature Photonics},
  title     = {Reversible optical control of macroscopic polarization in ferroelectrics},
  year      = {2017},
  issn      = {1749-4893},
  month     = dec,
  number    = {1},
  pages     = {29--32},
  volume    = {12},
  doi       = {10.1038/s41566-017-0068-1},
  publisher = {Springer Science and Business Media LLC},
}

@Article{Vats2019,
  author    = {Vats, Gaurav and Bai, Yang and Zhang, Dawei and Juuti, Jari and Seidel, Jan},
  journal   = {Advanced Optical Materials},
  title     = {Optical Control of Ferroelectric Domains: Nanoscale Insight into Macroscopic Observations},
  year      = {2019},
  issn      = {2195-1071},
  month     = mar,
  number    = {11},
  volume    = {7},
  doi       = {10.1002/adom.201800858},
  publisher = {Wiley},
}

@Article{Wang2019,
  author    = {Wang, Hua and Qian, Xiaofeng},
  journal   = {Science Advances},
  title     = {Ferroicity-driven nonlinear photocurrent switching in time-reversal invariant ferroic materials},
  year      = {2019},
  issn      = {2375-2548},
  month     = aug,
  number    = {8},
  volume    = {5},
  doi       = {10.1126/sciadv.aav9743},
  publisher = {American Association for the Advancement of Science (AAAS)},
}

@Article{RosePetruck1999,
  author    = {Rose-Petruck, Christoph and Jimenez, Ralph and Guo, Ting and Cavalleri, Andrea and Siders, Craig W. and Rksi, Ferenc and Squier, Jeff A. and Walker, Barry C. and Wilson, Kent R. and Barty, Christopher P. J.},
  journal   = {Nature},
  title     = {Picosecond–milliångström lattice dynamics measured by ultrafast X-ray diffraction},
  year      = {1999},
  issn      = {1476-4687},
  month     = mar,
  number    = {6725},
  pages     = {310--312},
  volume    = {398},
  doi       = {10.1038/18631},
  publisher = {Springer Science and Business Media LLC},
}

@Article{Whiteley2019,
  author    = {Whiteley, S. J. and Heremans, F. J. and Wolfowicz, G. and Awschalom, D. D. and Holt, M. V.},
  journal   = {Nature Communications},
  title     = {Correlating dynamic strain and photoluminescence of solid-state defects with stroboscopic x-ray diffraction microscopy},
  year      = {2019},
  issn      = {2041-1723},
  month     = jul,
  number    = {1},
  volume    = {10},
  doi       = {10.1038/s41467-019-11365-9},
  publisher = {Springer Science and Business Media LLC},
}

@Article{Mariette2021,
  author    = {Mariette, C. and Lorenc, M. and Cailleau, H. and Collet, E. and Guérin, L. and Volte, A. and Trzop, E. and Bertoni, R. and Dong, X. and Lépine, B. and Hernandez, O. and Janod, E. and Cario, L. and Ta Phuoc, V. and Ohkoshi, S. and Tokoro, H. and Patthey, L. and Babic, A. and Usov, I. and Ozerov, D. and Sala, L. and Ebner, S. and Böhler, P. and Keller, A. and Oggenfuss, A. and Zmofing, T. and Redford, S. and Vetter, S. and Follath, R. and Juranic, P. and Schreiber, A. and Beaud, P. and Esposito, V. and Deng, Y. and Ingold, G. and Chergui, M. and Mancini, G. F. and Mankowsky, R. and Svetina, C. and Zerdane, S. and Mozzanica, A. and Bosak, A. and Wulff, M. and Levantino, M. and Lemke, H. and Cammarata, M.},
  journal   = {Nature Communications},
  title     = {Strain wave pathway to semiconductor-to-metal transition revealed by time-resolved X-ray powder diffraction},
  year      = {2021},
  issn      = {2041-1723},
  month     = feb,
  number    = {1},
  volume    = {12},
  doi       = {10.1038/s41467-021-21316-y},
  publisher = {Springer Science and Business Media LLC},
}

@Article{Wang2022,
  author    = {Wang, Yingqi and Liu, Cunming and Ren, Yang and Zuo, Xiaobing and Canton, Sophie E. and Zheng, Kaibo and Lu, Kuangda and Lü, Xujie and Yang, Wenge and Zhang, Xiaoyi},
  journal   = {Journal of the American Chemical Society},
  title     = {Visualizing Light-Induced Microstrain and Phase Transition in Lead-Free Perovskites Using Time-Resolved X-Ray Diffraction},
  year      = {2022},
  issn      = {1520-5126},
  month     = mar,
  number    = {12},
  pages     = {5335--5341},
  volume    = {144},
  doi       = {10.1021/jacs.1c11747},
  publisher = {American Chemical Society (ACS)},
}

@Article{Slezak2020,
  author    = {Slezák, Ondřej and Lucianetti, Antonio and Mocek, Tomáš},
  journal   = {Journal of the Optical Society of America B},
  title     = {Tensor-to-matrix mapping in elasto-optics},
  year      = {2020},
  issn      = {1520-8540},
  month     = Mar,
  number    = {4},
  pages     = {1090},
  volume    = {37},
  doi       = {10.1364/josab.383975},
  publisher = {Optica Publishing Group},
}

@Article{Shugaev2021,
  author    = {Shugaev, Maxim V. and Zhigilei, Leonid V.},
  journal   = {Journal of Applied Physics},
  title     = {Thermoelastic modeling of laser-induced generation of strong surface acoustic waves},
  year      = {2021},
  issn      = {1089-7550},
  month     = Nov,
  number    = {18},
  volume    = {130},
  doi       = {10.1063/5.0071170},
  publisher = {AIP Publishing},
}

@Article{Zarei2023,
  author    = {Zarei, Alireza and Pilla, Srikanth},
  journal   = {International Journal of Heat and Mass Transfer},
  title     = {An improved theory of thermoelasticity for ultrafast heating of materials using short and ultrashort laser pulses},
  year      = {2023},
  issn      = {0017-9310},
  month     = Nov,
  pages     = {124510},
  volume    = {215},
  doi       = {10.1016/j.ijheatmasstransfer.2023.124510},
  publisher = {Elsevier BV},
}

@Article{Qian2016,
  author    = {Qian, Jing and Wang, Chengwei and Huang, Yuanyuan and Li, Hongjing and Lou, Kongyu and Wang, Guande and Zhao, Quan-Zhong},
  journal   = {Applied Optics},
  title     = {Ultrashort pulsed laser induced heating-nanoscale measurement of the internal temperature of dielectrics using black-body radiation},
  year      = {2016},
  issn      = {1539-4522},
  month     = Oct,
  number    = {29},
  pages     = {8347},
  volume    = {55},
  doi       = {10.1364/ao.55.008347},
  publisher = {Optica Publishing Group},
}

@InBook{Authier2013,
  author    = {Authier, A.},
  pages     = {3--33},
  publisher = {International Union of Crystallography},
  title     = {Introduction to the properties of tensors},
  year      = {2013},
  isbn      = {9781118762295},
  month     = Dec,
  booktitle = {International Tables for Crystallography},
  doi       = {10.1107/97809553602060000900},
}

@Article{WillettGies2014,
  author    = {Willett-Gies, Travis and DeLong, Eric and Zollner, Stefan},
  journal   = {Thin Solid Films},
  title     = {Vibrational properties of bulk \ch{LaAlO3} from Fourier-transform infrared ellipsometry},
  year      = {2014},
  issn      = {0040-6090},
  month     = nov,
  pages     = {620--624},
  volume    = {571},
  doi       = {10.1016/j.tsf.2013.11.140},
  publisher = {Elsevier BV},
}

@Article{Carpenter2009,
  author    = {Carpenter, M A and Sinogeikin, S V and Bass, J D and Lakshtanov, D L and Jacobsen, S D},
  journal   = {Journal of Physics: Condensed Matter},
  title     = {Elastic relaxations associated with the {$Pm\bar{3}m$}–{$R\bar{3}c$} transition in \ch{LaAlO3} I. Single crystal elastic moduli at room temperature},
  year      = {2009},
  issn      = {1361-648X},
  month     = Dec,
  number    = {3},
  pages     = {035403},
  volume    = {22},
  comment   = {Thermoelastic coefficients at room temp.},
  doi       = {10.1088/0953-8984/22/3/035403},
  publisher = {IOP Publishing},
}

@Article{VillasBoas2019,
  author    = {Villas-Boas, Lúcia and Goulart, Celso and Ferreira, de},
  journal   = {Processing and Application of Ceramics},
  title     = {Effects of Sr and Mn co-doping on microstructural evolution and electrical properties of \ch{LaAlO3}},
  year      = {2019},
  issn      = {2406-1034},
  number    = {4},
  pages     = {333--341},
  volume    = {13},
  doi       = {10.2298/pac1904333v},
  publisher = {National Library of Serbia},
}

@Article{Silva2015,
  author    = {da Silva, Cristiane A. and de Miranda, Paulo Emílio V.},
  journal   = {International Journal of Hydrogen Energy},
  title     = {Synthesis of \ch{LaAlO3} based materials for potential use as methane-fueled solid oxide fuel cell anodes},
  year      = {2015},
  issn      = {0360-3199},
  month     = Aug,
  number    = {32},
  pages     = {10002--10015},
  volume    = {40},
  comment   = {Thermal expansion coefficient},
  doi       = {10.1016/j.ijhydene.2015.06.019},
  publisher = {Elsevier BV},
}

\newpage
\appendix
\renewcommand{\appendixname}{Supplementary material}

\begin{appendices}
\onecolumngrid

\section{Effects of laser-induced heating on strain and birefringence}
Laser light travelling through a medium will deposit part of its energy into the material, thereby heating it. This causes the material to expand in the heated regions. The uneven heating by a pump-pulse causes non-uniform expansion, resulting in spatially varying strains. The temperature gradient subsequently creates a heat-flow and thus a temporal temperature and strain distribution. These effects can be described by a system of coupled thermoelastic equations in which the heat flow is dependent on both the temperature profile and the induced strain \cite{Zarei2023}. To simplify calculations, the response of our crystals to laser-induced heating was modelled with an uncoupled theory, in which it is assumed that temperature-induced strains have no meaningful effect on heat flow.

\subsection{Simulated heat flow}
Upon irradiation with a spatially Gaussian pump pulse the crystal is locally heated, with the temperature profile following the same spatial profile of the pump. The energy deposition throughout the depth of the crystal is assumed according to the Beer-Lambert law, with an exponential decay \cite{Shugaev2021}. The model assumes that the deposited energy is instantaneously converted into heat, although more sophisticated models like the Lord-Schulman and Green-Lindsay include relaxation times in their thermoelastic equations to account for the different energy pathways towards thermal, resulting in much lower (15 - 25\%) peak surface temperatures compared to the classical model \cite{Zarei2023}. Black body radiation measurements have been used to observe surface temperatures in dielectrics after irradiation with ultrafast (femtosecond) pulses, and show the temperature evolution is governed by shock waves in the first several nanoseconds, after which thermal diffusion becomes the dominant effect \cite{Qian2016}. To account for these effects, the effective energy delivered as heat into the sample can be taken as a fraction of the experimentally measured pulse energy.

The simulated crystal is divided into regularly spaced voxels with volume $V=h_x\times h_y\times h_z$, with an initial temperature given by

\begin{equation} \label{eq:voxelTemperature}
\begin{aligned}
    \Delta T(x, y, z) &= \frac{Q_{voxel}(x, y, z)}{\rho V C_p} \\
    &= \frac{Q_{pulse}}{\rho h_x C_p}\frac{1}{2\pi\sigma^2} \cdot
    e^{-\frac{y^2+z^2}{2\sigma^2} - \alpha x} \cdot (1 - e^{-\alpha h_x}),
\end{aligned}
\end{equation}
where $Q_{pulse}$ and $\sigma$ are the energy and standard deviation of the spatially Gaussian pump pulse, and $\rho$, $C_p$ and $\alpha$ the density, specific heat and absorption coefficient of the material.

The heat flow is subsequently numerically solved with the classical transient heat equation

\begin{equation}
    q_i = -k_{ij}\cdot \frac{d(\Delta T)}{dx_j},
\end{equation}
with $q$ the heat flux, $k$ the thermal conductivity and $\Delta T = T - T_0$ the spatial temperature profile with respect to a reference temperature $T_0$.

The absorption coefficient in the investigated spectral region ranges between \qtyrange{0}{8}{\um^{-1}} in \ch{LaAlO3}. This value greatly influences the maximum temperature that the crystal can reach, with $\Delta T \propto \alpha$ scaling with the absorption coefficient. Plugging realistic values into Eq. \ref{eq:voxelTemperature} of $Q_{pulse}=\qty{100}{\uJ}$ and $\sigma=\qty{60}{\um}$ yields a $\Delta T_{max}=\qty{73}{K}$ for $\alpha=\qty{0.05}{\um^{-1}}$ and $T_{max}=\qty{2937}{K}$ for $\alpha=\qty{2.0}{um^{-1}}$. It should be noted that higher absorption coefficients often coincide with high reflectivities, thus reducing the effective thermal energy increase substantially.

For surface temperatures up to several thousand Kelvin, convection and thermal radiation effects are several orders of magnitude lower than thermal conduction, and are thus not considered. Values of the relevant thermal material properties are shown in table \ref{tab:lao_thermal_properties}. 

\begin{table}[]
    \centering
    \caption{Measured quantities of \ch{LaAlO3} used in the thermoelastic simulations.}
    \renewcommand{\arraystretch}{1.2}
    \begin{tabular}{ cccccc }

        Quantity & Measured value & Units & Description \\ \hline\hline
        $\rho$ & \num{6.52} \cite{VillasBoas2019} & \unit{g.cm^{-3}} & Density \\
        $C_P$ & \num{0.427} \cite{Michael1992} & \unit{J.g^{-1}.K^{-1}}  & Specific heat\\
        $k$ & \num{13.6} \cite{Michael1992} & \unit{W.m^{-1}.K^{-1}} & Thermal conductivity \\
        $\alpha_{CTE}$ & \num{11.4} \cite{Silva2015} & \qty{d-6}{K^{-1}} & Thermal expansion coefficient\\
        \\
        \multicolumn{4}{c}{Thermoelastic coefficients \cite{Carpenter2009}} \\
        \hline
        $C_{11}$ & 337 & \unit{GPa} & \\
        $C_{12}$ & 151 & & \\
        $C_{13}$ & 93 & & \\
        $C_{14}$ & 46 & & \\
        $C_{33}$ & 411 & & \\
        $C_{44}$ & 121 & & \\
    \end{tabular}
    \label{tab:lao_thermal_properties}
\end{table}

\subsection{Thermoelastic strain}
The simulated regular grid of temperature nodes is used to compute the resulting strain by solving the linear system of equations

\begin{equation}
    K\cdot u = f,
\end{equation}
where $K$ is the global stiffness matrix, $f$ the thermal load vector and $u$ the displacement vector of every temperature node. The global stiffness matrix and thermal load vector are assembled by dividing the temperature grid in so-called elements, which consist of a group of temperature nodes spanning a volume $V$. For each of these elements, the elemental stiffness matrix $K_e$ and load vector $f_e$ can be defined as

\begin{align}
    K_e &= \int_V B^T C^E B dV, \label{eq:elementStiffnessMatrix} \\
    f_e &= \int_V B^T C^E \varepsilon_{th} dV, \label{eq:elementThermalLoadVector}
\end{align}
where $B$ is the strain-displacement matrix, $C^E$ the elastic stiffness matrix at constant electric field and $\varepsilon_{th} = \alpha_{CTE}\Delta T$ the thermal strain. The elastic stiffness matrix is highly dependent on crystal symmetry and has a large influence on the resulting strains. The matricex for point group $\bar{3}$m (\ch{LaAlO3} at room temperature) is given in Eq \eqref{eq:LAOElasticStiffness} \cite{Authier2013}.

\begin{equation}\label{eq:LAOElasticStiffness}
    C^E = 
    \begin{pmatrix}
        C_{11} & C_{12} & C_{13} & C_{14} &      0 &      0 \\
        C_{12} & C_{11} & C_{13} &-C_{14} &      0 &      0 \\
        C_{13} & C_{13} & C_{33} &      0 &      0 &      0 \\
        C_{14} &-C_{14} &      0 & C_{44} &      0 &      0 \\
             0 &      0 &      0 &      0 & C_{44} & C_{14} \\
             0 &      0 &      0 &      0 & C_{14} & \frac{C_{11} - C_{12}}{2}
    \end{pmatrix}
\end{equation}

Due to the regularity of the temperature grid, an 8-node trilinear hexahedron (HEX8) was chosen as the element shape. The strain-displacement matrix for such HEX8 elements is given by the $(6\times24)$ matrix

\begin{align}
    B &= \left[
        B_1,\ldots,B_8
    \right]\mathrm{, where} \\
    B_i &= \left[
    \begin{matrix}
        \frac{\partial N_i^e}{\partial x} & 0 & 0 \\
        0 & \frac{\partial N_i^e}{\partial y} & 0 \\
        0 & 0 & \frac{\partial N_i^e}{\partial z} \\
        \frac{\partial N_i^e}{\partial y} & \frac{\partial N_i^e}{\partial x} & 0 \\
        0 & \frac{\partial N_i^e}{\partial z} & \frac{\partial N_i^e}{\partial y} \\
        \frac{\partial N_i^e}{\partial z} & 0 & \frac{\partial N_i^e}{\partial x}
    \end{matrix}
    \right]\mathrm{, and} \\
    N_i^e(\xi, \eta, \zeta) & = \frac{1}{8}(1+\xi\xi_i)(1+\eta\eta_i)(1+\zeta\zeta_i)
\end{align}
where $N_i^e$ are the shape functions that define the displacement within an element by linear approximation, 

\begin{equation}
    \begin{aligned}
        u_x(x, y, z) &\approx \sum_{i=1}^{8} N_i^e (u_x)_i \\
        u_y(x, y, z) &\approx \sum_{i=1}^{8} N_i^e (u_y)_i \\
        u_z(x, y, z) &\approx \sum_{i=1}^{8} N_i^e (u_z)_i \\
    \end{aligned}
\end{equation}
with $u_x, u_y, u_z$ the displacement in the $x$, $y$ and $z$ direction respectively as a function of the displacement of the nodes spanning the element. To simplify the maths, a coordinate transformation to the natural coordinate system $(\xi, \eta, \zeta)$ is made where $\xi_i$, $\eta_i$ and $\zeta_i$ are the natural coordinates of the $i^{th}$ node. In this coordinate system, every node is placed at one of the coordinates $\xi_i, \eta_i, \zeta_i \in \{-1, 1\}$. The regularity of the HEX8 system simplifies the integral in Eq. \ref{eq:elementStiffnessMatrix} to a summation with $B_m \equiv B(\xi_m, \eta_m, \zeta_m)$ evaluated at the eight so-called Gauss points $G_m=(\xi_m, \eta_m, \zeta_m)\ |\ m=\{1,\ldots,8\}$ where $\xi_m, \eta_m, \zeta_m \in \{-\frac{1}{\sqrt{3}}, \frac{1}{\sqrt{3}}\}$. Subscript $m$ is thus taken to mean the function is evaluated at coordinate $G_m$.

\begin{equation}
    K_e = \frac{h_xh_yh_z}{8} \sum_{m=1}^{8} B_m^T C^E B_m.
\end{equation}

\begin{figure}
    \centering
    \includegraphics[width=5in]{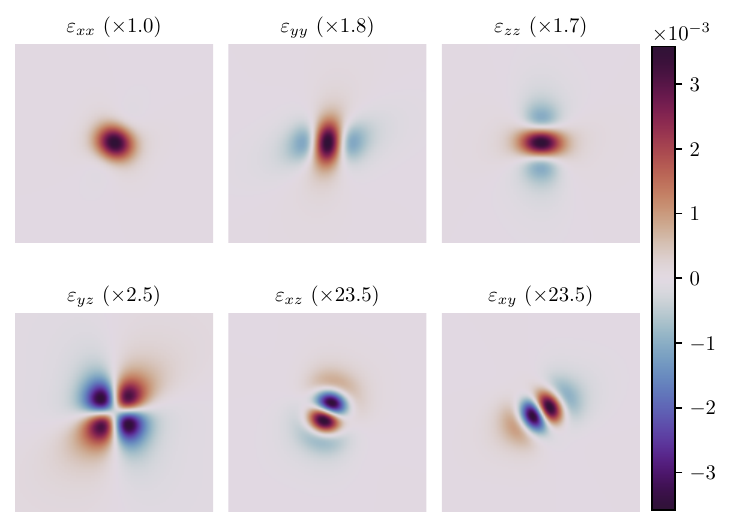}
    \caption{Simulated strain at the (\qty{1}{mm^{2}}) $yz$-surface of a \ch{LaAlO3} crystal \qty{100}{\us} after excitation with a \qty{1}{mJ} Gaussian pulse with a FWHM of \qty{110}{\um}, and absorption coefficient $\alpha=\qty{0.03}{\um^{-1}}$.}
    \label{fig:LAO_2D_calcStrain}
\end{figure}

\begin{figure}
    \centering
    \includegraphics[width=4in]{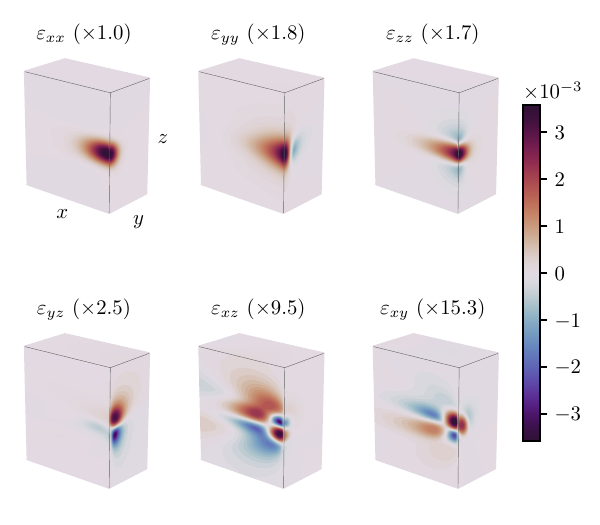}
    \caption{3D cut of the simulated strain described in Fig. \ref{fig:LAO_2D_calcStrain} to a depth of $x=\qty{250}{\um}$.}
    \label{fig:LAO_3D_calcStrain}
\end{figure}

The load vector is simplified similarly to

\begin{equation}
    f_e = \frac{h_xh_yh_z}{8} \sum_{m=1}^{8}\sum_{i=1}^{8} \left(N_i^e\right)_m B_m^T C^E \alpha_{CTE} \Delta T_i.
\end{equation}
From these elements, the global stiffness matrix and load vector are constructed. Free boundary conditions are used with 3-2-1 constraints to remove the translational and rotational degrees of freedom from the solution by fixing one node in every direction, a second in two directions, and a third in one direction. The system is subsequently solved for the displacement $u$. With the displacement, the strain and stress can be calculated as

\begin{equation}
    \begin{aligned}
        \varepsilon_e &= B(0,0,0) u_e \\
        \sigma_e &= C^E \left(\varepsilon_e - (\varepsilon_{th})_e \right)
    \end{aligned}
\end{equation}
with the strain and stress both evaluated at the center of each element ($\xi=\eta=\zeta=0$). The resulting strains are shown in Figs \ref{fig:LAO_2D_calcStrain} and \ref{fig:LAO_3D_calcStrain}.

The asymmetry in the strain components of \ch{BaTiO3} is largely caused by the negative thermal expansion coefficient along the polarization axis in the tetragonal phase of \ch{BaTiO3}, while the anisotropic elastic stiffness tensor is the main cause of the asymmetry in the \ch{LaAlO3} strain terms.

\section{Millisecond domain switching}
Temporal domain switching and strain in \ch{LaAlO3} were measured on the millisecond timescale after pumping the crystal with a 12 µm macropulse. This macropulse consists of a 10-µs-long train of picosecond pulses (micropulses) with a repetition rate of 25 MHz, and a total power of 1 mJ. The macropulse can generate a much larger strain and switching than single micropulses, at the cost of picosecond temporal resolution. Here, it is used to study the millisecond decay of both the strain pattern and induced switching. Both strain and switching were measured with polarization sensitive microscopy, using a 515 nm CW laser as a probe. Images were captured on a CMOS camera with a 27 µs exposure time.

\begin{figure}
    \centering
    \includegraphics[width=3.in]{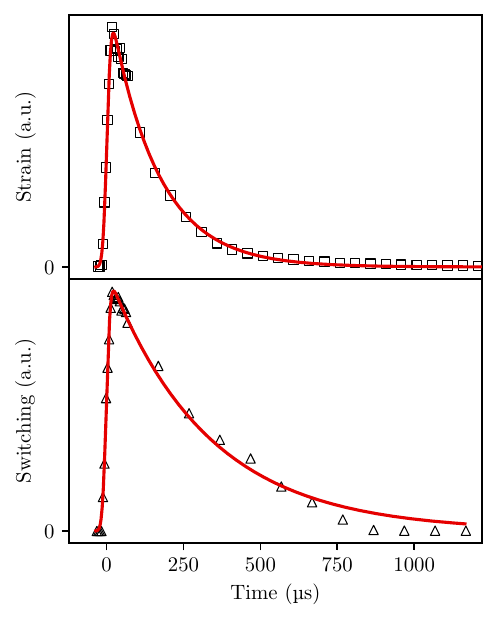}
    \caption{Induced strain and domain switching after pumping with a single micropulse at 12 µm. The strain graph shows the magnitude of the quasi-static strain, averaged over an area of $1\times1$ µm$^2$. The switching graph shows the total area of switched domains. The red lines show the fit of an exponentially modified Gaussian distribution.}
    \label{fig:ms-time-dependence}
\end{figure}

Both strain and domains appear within the time resolution of this experimental setup (30 µs) before decaying on a millisecond timescale (Fig. \ref{fig:ms-time-dependence}). A fit of an exponentially modified gaussian (EMG) was made for both the strain and domain switching, giving their characteristic lifetimes of $\tau=154$ µs and $\tau=324$ µs respectively. In contrast, temporal measurements with picosecond pulses as a pump show a faster decay for domains than for strain (as described in the main text). 
This can be explained by the macropulse creating a much larger strain profile than a micropulse, both in area and amplitude. Thus, stable switching can occur in a larger area, resulting in much larger domains. These larger domain sizes are more thermodynamically stable, resulting in a longer lifetime, even when the strain has already dissipated. The interplay between strain and domain lifetime offers an explanation as well for the worse fit of the EMG to the switching data.

\section{Infrared spectrum of domain switching}
\begin{figure}
    \centering
    \includegraphics[width=3in]{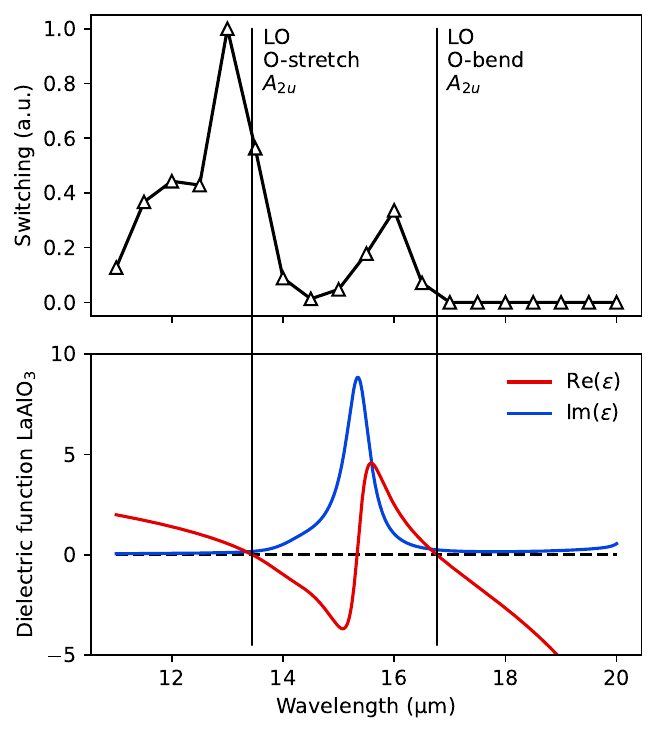}
    \caption{Domain switching as a function of pump wavelength. The real and imaginary part of the dielectric function show the LO frequencies in this spectral range \cite{WillettGies2014}. Switching peaks close to the epsilon-near-zero (ENZ) points.}
    \label{fig:spectrum-switching}
\end{figure}

A spectral dependence on domain switching in \ch{LaAlO3} was measured to find the optimal wavelength for studying optical domain switching and its dynamics. Previous research suggests domain switching in the ferroelectric \ch{BaTiO3} peaks near the epsilon-near-zero (ENZ) wavelengths \cite{Kwaaitaal2024}. Figure \ref{fig:spectrum-switching} shows a very similar result, with domain switching peaking at slightly blue-shifted wavelengths from the LO phonons. The dielectric function in this figure was calculated with experimental data from Willet-Gies et al. \cite{WillettGies2014}. From this spectrum, a wavelength range of 12-14 µm was chosen as the pump wavelength in our study.

\end{appendices}

\end{document}